\documentclass[12pt,onecolumn,draft]{IEEEtran} 

\usepackage{amsmath,amssymb,amsfonts}
\usepackage[final]{graphicx}
\usepackage{psfrag}
\usepackage[numbers,sort&compress]{natbib}

\newtheorem{corollary}{Corollary}

\newtheorem{proposition}{Proposition}
\newtheorem{lemma}{Lemma}
\newtheorem{definition}{Definition}

\def \H {{\mathcal H }}
\def \cov {\textit{cov}}

\begin{document}
\title{\Large{On the Bivariate Nakagami-$m$ Cumulative Distribution Function: Closed-form Expression and Applications}}

\author{\normalsize{F.J. Lopez-Martinez, D. Morales-Jimenez, E. Martos-Naya, J.F. Paris}\thanks{F.J. Lopez-Martinez, E. Martos-Naya and J.F. Paris are with Dept. Ingenieria de Comunicaciones at University of Malaga (Spain). D. Morales-Jimenez is with Dept. Information and Communication Technologies at Universitat Pompeu Fabra (Spain).}
\thanks{This work has been submitted to the IEEE for possible publication. Copyright may be transferred without notice, after which this version may no longer be accessible.}}
\maketitle

\begin{abstract}
In this paper, we derive exact closed-form expressions for the bivariate Nakagami-$m$ cumulative distribution function (CDF) with positive integer fading severity index $m$ in terms of a class of hypergeometric functions. Particularly, we show that the bivariate Nakagami-$m$ CDF can be expressed as a finite sum of elementary functions and bivariate confluent hypergeometric $\Phi_3$ functions. Direct applications which arise from the proposed closed-form expression are the outage probability (OP) analysis of a dual-branch selection combiner in correlated Nakagami-$m$ fading, or the calculation of the level crossing rate (LCR) and average fade duration (AFD) of a sampled Nakagami-$m$ fading envelope. 
\end{abstract}

\begin{IEEEkeywords}
Average fade duration, Bivariate Nakagami-$m$, Correlation, Cumulative distribution function, Diversity reception, Level crossing rate, Nakagami-$m$ fading.
\end{IEEEkeywords}

\section{Introduction}
There exist a number of statistical distributions which model the random fluctuations of the signal amplitude when transmitted through a wireless channel \cite{Simon2005}. Besides Rayleigh, Rician and other fading models, the Nakagami-$m$ distribution \cite{Nakagami1960} is an empirically-validated distribution which is extensively used to model different propagation conditions by means of changing the value of the fading severity index $m$. This distribution includes Rayleigh and one-sided Gaussian as particular cases, and approximates very closely both Rician and lognormal distributions \cite{Filho2004}.

While the PDF and the CDF of a Nakagami-$m$ variate follow very simple and tractable expressions, the statistical characterization of the multivariate Nakagami-$m$ distribution is rather involved and remains as an open problem in the literature. Although a closed-form expression for the joint PDF of two correlated Nakagami-$m$ variates was given in the original manuscript by Nakagami \cite{Nakagami1960}, a closed-form expression for the bivariate Nakagami-m CDF has not been obtained yet to the best of our knowledge. Existing results provide expressions for this bivariate CDF in terms of infinite summations \cite{Tan1997,Reig2002,deSouza2008}, and more recently \cite{Beaulieu2011} in terms of a single integral involving special functions. The result in \cite{Beaulieu2011} is particularly relevant, since it allows to express the PDF and the CDF of a number of multivariate distributions in terms of a single integral, whilst previous results were only available in terms on nested infinite summations \cite{Dharmawansa2007,Peppas2009}.

As the Rayleigh distribution is coincident with the Nakagami-$m$ distribution when the fading parameter $m=1$, it may result expectable that the existing closed-form expression for the bivariate Rayleigh CDF \cite{Schwartz1966} in terms of the first order Marcum-$Q$ function is a particular case of the bivariate Nakagami-$m$ CDF. This aspect was discussed by Simon and Alouini in \cite[p. 174]{Simon2005}, who pointed out that \textit{``One might anticipate that the bivariate Nakagami-m CDF could be expressed in a form analogous to (the bivariate Rayleigh CDF) depending instead on the $m^{th}$-order Marcum $Q$-function ... ... Unfortunately, to the author's knowledge an expression analogous to (the bivariate Rayleigh CDF) has not been reported in the literature, and the authors have themselves been unable to arrive at one.''}.

In this paper, we demonstrate that the bivariate Nakagami-$m$ cumulative distribution function can be expressed in closed-form as a finite sum of elementary functions and $\Phi_3$ functions, for a positive integer fading severity index $m$. The confluent hypergeometric function of two variables $\Phi_3$, which is well studied in classical books of integrals and special functions \cite[9.261.3]{Gradstein2007} and Laplace transforms \cite{Erdelyi1954}, is one of the bivariate forms of Kummer's confluent hypergeometric function $_1F_1$ and is often referred to as the Horn function \cite{Erdelyi1953}. Interestingly, there exists a connection between this kind of hypergeometric functions and the Marcum-$Q$ function \cite{Brychkov2012} of first-order, which allows to reduce the proposed expression for the bivariate CDF to the Rayleigh case when $m=1$. Hence, the connection foreseen by Simon and Alouini is found at the level of this family of hypergeometric functions, and a further connection between $\Phi_3$ and generalized Marcum-$Q$ functions of higher orders may be inferred.

The bivariate confluent hypergeometric function $\Phi_3$ appears in the literature in a number of scenarios, e.g. in the CDF of the minimum eigenvalue of a correlated non-central Wishart matrix \cite{Dharmawansa2011}, or in statistical models for multisensor synthetic aperture radar images \cite{Chatelain2008}. Although $\Phi_3$ function is not yet included in most commercial mathematical packages, its calculation can be easily performed using a numerical Laplace transform inversion \cite{Simon2005,Abate1995}. Therefore, a simple Mathematica$\texttrademark$ routine for the calculation of the confluent hypergeometric function $\Phi_3$ is given. This program provides accurate results for any values of the power correlation coefficient ${0<\rho<1}$, thus circumventing some of the issues related with the convergence of the infinite series expressions for the CDF \cite{Tan1997,Reig2002,deSouza2008} as ${\rho \to 1}$.

Summarizing, the main mathematical contributions of this paper are listed below

\begin{itemize}
\item{A closed-form expression for the bivariate Nakagami-$m$ CDF is given, in terms of a finite sum of elementary functions and bivariate confluent hypergeometric functions $\Phi_3$.}
\item{The connection between the closed-form expression for the bivariate Rayleigh CDF \cite{Schwartz1966} and the derived CDF expression with $m=1$ is established.}
\end{itemize}

These relevant results find numerous applications in the field of wireless communications; particularly:

(1) We provide a closed-form expression for the outage probability (OP) of a dual branch selection combining (SC) scheme under correlated Nakagami-$m$ fading.

(2) We derive closed-form expressions for the level crossing rate (LCR) and the average fade duration (AFD) of a sampled Nakagami-$m$ fading envelope, following the new approach introduced in \cite{Lopez2012}.

The remainder of this paper is organized as follows: in Section \ref{Analysis}, the main mathematical contributions of this paper are presented. Section \ref{Applications} is dedicated to analyze some scenarios of interest in communications making use of the derived expressions for the Nakagami-$m$ bivariate CDF. Numerical results are introduced in Section \ref{Numerical}, whereas the main conclusions are exposed in Section \ref{Conclusion}.

\section{Statistical Analysis} 
\label{Analysis}

\subsection{Notation and preliminaries}

In the following, we use $|\cdot|$ to indicate the modulus of a complex number, and $\text{E}\{\cdot\}$ to represent the expectation operation. The correlation coefficient of two random variables $X$ and $Y$ is defined as
\begin{equation}
\label{powcorr}
{\rho_{X,Y}\triangleq \frac{\cov\{X,Y\}}{\sqrt{\sigma_{X}^2{\sigma_{Y}^2}}}},
\end{equation}
where $\cov\{\}$ denotes covariance operation, and $\sigma_{X}^2$,$\sigma_{Y}^2$ represent the variance of the random variables $X$ and $Y$ respectively.

The CDF of a random variable $X$ is defined as $F_X(x)\triangleq \Pr\{X\leq x\}$, and consequently, the complementary CDF (CCDF) of $X$ is defined as $\bar{F}_X(x)\triangleq \Pr\{X > x\}=1-F_X(x).$

The joint CDF of two correlated random variables $X$ and $Y$ is defined as
\begin{equation}
{F_{X,Y}(x,y)}\triangleq \Pr\{X\leq x,Y\leq y\},
\end{equation}
whereas their joint CCDF is defined as
\begin{equation}
\label{relCDF}
{\bar{F}_{X,Y}(x,y)}\triangleq \Pr\{X> x,Y > y\}=F_{X,Y}(x,y)-F_X(x)-F_Y(y)+1.
\end{equation}


For the sake of compactness, we use the notation $\oint_{\Gamma}\triangleq \frac{1}{2\pi j}\int_{\Gamma}$, where $\int_{\Gamma}$ represents the integral along a contour $\Gamma$.

We introduce an integral associated with the Marcum-$Q$ function of $m$-th order, as well as some contour integrals of interest.
\begin{definition}
\textit{Incomplete integral $\H_m\left(u,\gamma,\delta\right)$ of $Q_m$ function.}
\label{def1}
\begin{equation}
\label{eqdef1}
\H_m\left(u,\gamma,\delta\right) \triangleq {\int_u^\infty  {x^{2m - 1} \exp \left( { - m x^2 } \right)Q_m \left( {\gamma x,\delta } \right)dx}},
\end{equation}
where $Q_m(a,b)$ is the $m^{th}$ order Marcum $Q$ function, $m \in \mathbb{N}, u \in [0,\infty)$ and $\delta,\gamma \in \mathbb{R}$.
\end{definition}

\begin{definition}
\textit{Contour integral $\mathcal{I}_k$}
\begin{equation}
\label{def2}
\mathcal{I}_k\left(u,\gamma,\delta\right)\triangleq \oint_\Gamma  {\frac{1}{{p - 1}}p^{ - k} \exp \left( {\frac{{u^2 }}{2}\frac{{\gamma ^2 }}{p}} \right)\exp \left( {\frac{{\delta ^2 p}}{2}} \right)dp} 
\end{equation}
where $\Gamma$ is a circular contour of radius less than unity that encloses the origin, $k \in \mathbb{N}, u \in [0,\infty)$ and $\delta,\gamma \in \mathbb{R}$.
\end{definition}

\begin{definition}
\textit{Contour integral $\mathcal{J}_{i,k}$}
\begin{equation}
\label{def3}
\mathcal{J}_{i,k}\left(u,\gamma,\delta,\rho\right)\triangleq {\oint_\Gamma  {\frac{{1}}{{\left( {p - \rho } \right)^i }}p^{ - k} \exp \left( {\frac{{u^2 }}{2}\frac{{\gamma ^2 }}{p}} \right)\exp \left( {\frac{{\delta ^2 p}}{2}} \right)dp} }
\end{equation}
where $\Gamma$ is a circular contour of radius less than unity and greater than $\rho$ that encloses the origin, $\{i,k\} \in \mathbb{N}, \rho \in (0,1), u \in [0,\infty)$ and $\delta,\gamma \in \mathbb{R}$.
\end{definition}

\subsection{Results}

In this subsection, the main mathematical contributions of this paper are presented. In order to obtain a closed-form expression for the integral $\mathcal{H}_m$, we first establish the connection with the integrals $\mathcal{I}_k$ and $\mathcal{J}_{i,k}$.

\begin{proposition}
\label{prop3}
The integral $\H_m\left(u,\gamma,\delta\right)$ can be expressed as
\begin{equation}
\label{eqprop3}
\H_m\left(u,\gamma,\delta\right)=\exp \left( { - \frac{{\delta ^2 +\alpha u^2 }}{2}} \right)2^{m - 1} \alpha ^{ - m} \left( {m - 1} \right)!\sum\limits_{k = 0}^{m - 1} {\left( {\alpha \frac{{u^2 }}{2}} \right)^k \frac{1}{{k!}}} I_{m,k} \left( {u,\delta ,\gamma } \right),
\end{equation}
where $\alpha=\gamma^2+2m$ and
\begin{equation}
I_{m,k}= - \left( {1+\frac{\gamma^2}{2m }} \right)^{(m - k)} \mathcal{I}_k \left(u,\delta,\gamma\right)+ \sum\limits_{i = 1}^{m - k} {\left( {1+\frac{\gamma^2}{2m }} \right)^{(m - k - i +1)} \mathcal{J}_{i,k}\left(u,\delta,\gamma,\tfrac{\gamma^2}{\gamma^2+2m}\right)}.
\end{equation}
\end{proposition}

\begin{IEEEproof}
See Appendix \ref{App3}.
\end{IEEEproof}

Next, we express the integrals $\mathcal{I}_k$ and $\mathcal{J}_{i,k}$ in terms of the confluent hypergeometric $\Phi_3$ function.

\begin{proposition}
\label{prop1}
The contour integral $\mathcal{I}_k$ can be expressed in closed-form as
\begin{equation}
\label{eqprop1}
\mathcal{I}_k\left(u,\gamma,\delta\right)=\frac{{1}}{{\Gamma (k + 1)}}\left( {\frac{{\delta ^2 }}{2}} \right)^k \Phi _3 \left( {1,k + 1;\frac{{\delta ^2 }}{2},\frac{{u^2 }}{2}\frac{{\gamma ^2 \delta ^2 }}{2}} \right) - \exp \left( {\frac{{\gamma ^2 u^2 }}{2} + \frac{{\delta ^2 }}{2}} \right),
\end{equation}
where $\Phi_3\left(a,b;x,y\right)$ is the confluent hypergeometric function of two variables \cite[9.261.3]{Gradstein2007}.
\end{proposition}

\begin{IEEEproof}
See Appendix \ref{App1}.
\end{IEEEproof}

\begin{proposition}
\label{prop2}
The contour integral $\mathcal{J}_{i,k}$ can be expressed in closed-form as
\begin{equation}
\label{eqprop2}
\mathcal{J}_{i,k}\left(u,\gamma,\delta,\rho\right)=\frac{{1}}{{\Gamma (k + i)}}\left( {\frac{{\delta ^2 }}{2}} \right)^{k + i - 1} \Phi _3 \left( {i,k + i;\rho \frac{{\delta ^2 }}{2},\frac{{u^2 }}{2}\frac{{\gamma ^2 \delta ^2 }}{2}} \right),
\end{equation}
where $\Phi_3\left(a,b;x,y\right)$ is the confluent hypergeometric function of two variables.
\end{proposition}

\begin{IEEEproof}
See Appendix \ref{App2}.
\end{IEEEproof}

Note that combination of (\ref{eqprop1}), (\ref{eqprop2}) and (\ref{eqprop3}) gives the integral $\mathcal{H}_m$ in closed-form in terms of the $\Phi_3$ function.

\begin{lemma}
\label{lemma1}
The joint CCDF of two correlated Nakagami-$m$ random variables $R_1,R_2$ with positive integer fading severity index $m$, power correlation coefficient $\rho$, and $\Omega_1=\text{E}\left\{R_1^2\right\}$, $\Omega_2=\text{E}\left\{R_2^2\right\}$ power can be expressed as
\begin{equation}
\label{eqCCDF00}
\bar{F}_{R_1 ,R_2 } \left( {r_1,r_2;m,\rho } \right) =\frac{2m^m}{\Gamma(m)}\H_m\left(\frac{r_1}{\sqrt{\Omega_1}},\sqrt{\frac{2m\rho}{1-\rho}},\frac{r_2}{\sqrt{\Omega_2}}\sqrt{\frac{2m}{1-\rho}} \right)
\end{equation}
\end{lemma}

\begin{IEEEproof}
See Appendix \ref{App4}.
\end{IEEEproof}

\begin{corollary}
\label{corollary0}
Using the results in Lemma \ref{lemma1} and Propositions \ref{prop3}-\ref{prop2}, the joint CCDF of two correlated Nakagami-$m$ random variables $R_1,R_2$ with positive integer fading severity index $m$, power correlation coefficient $\rho$, and $\Omega_1=\text{E}\left\{R_1^2\right\}$, $\Omega_2=\text{E}\left\{R_2^2\right\}$ power can be expressed in closed-form as
\begin{align}
\label{eqCCDF01}
\bar{F}_{R_1 ,R_2 } \left( {r_1,r_2;m,\rho } \right) &= \sum\limits_{k = 0}^{m - 1} \left[ \exp \left( { - m\frac{r_1^2}{\Omega_1} } \right) - \left( {1 - \rho } \right)^{ - k} \exp \left( { - \frac{m}{{1 - \rho }}\left( {\frac{r_1^2}{\Omega_1}  + \frac{r_2^2}{\Omega_2} } \right)} \right) \right. \times \nonumber \\
&\left\{ \left( \frac{m r_2^2 }{\Omega_2}\right)^k\frac{1}{{\Gamma \left( {k + 1} \right)}}\Phi _3 \left( {1,k + 1;\tfrac{{r_2^2 }}{{\Omega_2}}\tfrac{{m}}{{(1 - \rho )}},\rho \left( \tfrac{{r_1}}{\sqrt{\Omega_1 }}\tfrac{{r_2}}{\sqrt{\Omega_2} }\tfrac{{m}}{(1 - \rho) } \right)^2 } \right) \right. - \nonumber \\ & \left.\left. \sum\limits_{i = 1}^{m - k} \left( \frac{m r_2^2 } {\Omega_2}\right)^{k+i-1}\frac{1}{{\Gamma \left( {k + i} \right)}}\Phi _3 \left( {i,k + i;\tfrac{{r_2^2 }}{{\Omega_2}}\tfrac{{m\rho}}{{(1 - \rho )}},\rho \left( \tfrac{{r_1}}{\sqrt{\Omega_1} }\tfrac{{r_2}}{\sqrt{\Omega_2} }\tfrac{{m}}{(1 - \rho) } \right)^2} \right)\right\} \right].
\end{align}
\end{corollary}

\begin{corollary}
\label{corollary1}
The joint CDF of two correlated Nakagami-$m$ random variables $R_1,R_2$ with positive integer fading severity index $m$, power correlation coefficient $\rho$, and $\Omega_1=\text{E}\left\{R_1^2\right\}$, $\Omega_2=\text{E}\left\{R_2^2\right\}$ power can be expressed in closed-form as
\begin{align}
\label{CDF01}
F_{R_1 ,R_2 } \left( {r_1,r_2;m,\rho } \right) &= 1- \sum\limits_{k = 0}^{m - 1} \left[{\exp \left( { - \frac{{mr_1 ^2 }}{{\Omega _1 }}} \right)\left[ {\left( {\frac{{mr_1 ^2 }}{{\Omega _1 }}} \right)^k \frac{1}{{k!}} - 1} \right] + \exp \left( { - \frac{{mr_2 ^2 }}{{\Omega _2 }}} \right)\left( {\frac{{mr_2 ^2 }}{{\Omega _2 }}} \right)^k \frac{1}{{k!}}} \right. \nonumber \\
 &- \left( {1 - \rho } \right)^{ - k} \exp \left( { - \frac{m}{{1 - \rho }}\left( {\frac{r_1^2}{\Omega_1}  + \frac{r_2^2}{\Omega_2} } \right)} \right)  \times \nonumber \\ \nonumber
&\left\{ \left( \frac{m r_2^2 }{\Omega_2}\right)^k\frac{1}{{\Gamma \left( {k + 1} \right)}}\Phi _3 \left( {1,k + 1;\tfrac{{r_2^2 }}{{\Omega_2}}\tfrac{{m}}{{(1 - \rho )}},\rho \left( \tfrac{{r_1}}{\sqrt{\Omega_1 }}\tfrac{{r_2}}{\sqrt{\Omega_2} }\tfrac{{m}}{(1 - \rho) } \right)^2 } \right) \right. - \\ & \left.\left. \sum\limits_{i = 1}^{m - k} \left( \frac{m r_2^2 } {\Omega_2}\right)^{k+i-1}\frac{1}{{\Gamma \left( {k + i} \right)}}\Phi _3 \left( {i,k + i;\tfrac{{r_2^2 }}{{\Omega_2}}\tfrac{{m\rho}}{{(1 - \rho )}},\rho \left( \tfrac{{r_1}}{\sqrt{\Omega_1} }\tfrac{{r_2}}{\sqrt{\Omega_2} }\tfrac{{m}}{(1 - \rho) } \right)^2} \right)\right\} \right].
\end{align}
\end{corollary}

\begin{IEEEproof}
The CDF of a Nakagami-$m$ variate is given by
\begin{equation}
\label{NakCDF}
F_X(x)=1-\frac{\Gamma\left(m,\tfrac{m}{\Omega} x^2\right)}{\Gamma(m)},
\end{equation}
where $\Gamma(\cdot,\cdot)$ is the upper incomplete Gamma function, and $\Gamma(\cdot)$ is the Gamma function. In case $m$ is a positive integer, (\ref{NakCDF}) can be expressed as

\begin{equation}
\label{NakCDF2}
F_X(x)=1-e^{-\frac{mx^2}{\Omega}}\sum_{k=0}^{m-1}\left(\frac{mx^2}{\Omega}\right)^k\frac{1}{k!}.
\end{equation}

Finally, using (\ref{NakCDF2}) and (\ref{eqCCDF01}) in (\ref{relCDF}), it is obtained the closed-form expression for the bivariate Nakagami-$m$ CDF in (\ref{CDF01}).
\end{IEEEproof}

These expressions for the bivariate Nakagami-$m$ CCDF (and thus for the CDF itself) are new to the best of our knowledge. These results are particularly relevant since no closed-form expressions were available for these functions in the literature. Expression (\ref{CDF01}) for the bivariate Nakagami-$m$ CDF is given in terms of a finite sum of $\Phi_3$ functions, which are listed in classical books of integrals and special functions \cite[9.261.3]{Gradstein2007}, as well as in the Laplace transforms tables \cite{Erdelyi1954}. Besides, it allows for connecting the results given in this paper with the existing results for the bivariate Rayleigh CDF thanks to the connection between $\Phi_3$ and $Q_1$ functions \cite{Brychkov2012}, as follows.

\begin{corollary}
\label{corollary2}
The joint CDF of two correlated Nakagami-$m$ random variables $R_1,R_2$ with fading severity index $m$, $\Omega_1=\text{E}\left\{R_1^2\right\}$, $\Omega_2=\text{E}\left\{R_2^2\right\}$ power, and power correlation coefficient $\rho$, specialized at $m=1$, reduces to the well-known expression for the bivariate Rayleigh CDF \cite{Schwartz1966} as
\begin{align}
\label{CDF2}
F_{R_1 ,R_2 } \left( {r_1,r_2;1,\rho } \right) &= 1 - \exp \left( { - \frac{{r_1^2 }}{{\Omega _1 }}} \right)Q_1 \left( {\sqrt {\frac{2}{{1 - \rho }}} \frac{{r_2 }}{{\sqrt {\Omega _2 } }},\sqrt {\frac{{2\rho }}{{1 - \rho }}} \frac{{r_1 }}{{\sqrt {\Omega _1 } }}} \right) \nonumber\\
&-\exp \left( { - \frac{{r_2^2 }}{{\Omega _2 }}} \right)\left[ {1 - Q_1 \left( {\sqrt {\frac{{2\rho }}{{1 - \rho }}} \frac{{r_2 }}{{\sqrt {\Omega _2 } }},\sqrt {\frac{2}{{1 - \rho }}} \frac{{r_1 }}{{\sqrt {\Omega _1 } }}} \right)} \right],
\end{align}
where $Q_1(a,b)$ is the first-order Marcum-$Q$ function.
\end{corollary}

\begin{IEEEproof}
Setting $m=1$ in (\ref{CDF01}), and defining $u=r_1/\sqrt{\Omega_1}$ and $v=r_2/\sqrt{\Omega_2}$ we have

\begin{align}
F_{R_1 ,R_2 } \left( {u,v;1,\rho } \right) &= \exp \left( { - u^2 } \right) - \left( {1 - \rho } \right)^{ - k} \exp \left( { - \frac{1}{{1 - \rho }}\left( {u^2  + v^2 } \right)} \right)\times \nonumber \\
&\left( {\Phi _3 \left( {1,1;\frac{{v^2 }}{{1 - \rho }},\rho \left( {\frac{{uv}}{{1 - \rho }}} \right)^2 } \right) - \Phi _3 \left( {1,1;\rho \frac{{v^2 }}{{1 - \rho }},\rho \left( {\frac{{uv}}{{1 - \rho }}} \right)^2 } \right)} \right).
\end{align}

Using the relationship between the confluent hypergeometric function $\Phi_3$ and the first-order Marcum-$Q$ function given in \cite{Brychkov2012}
\begin{align}
Q_1\left(a,b\right)=\exp\left(-\left(\frac{a^2}{2}+\frac{b^2}{2}\right)\right)\Phi_3\left(1,1;\frac{a^2}{2},\frac{a^2b^2}{4}\right),
\end{align}
and after some straightforward manipulations, it is obtained (\ref{CDF2}).
\end{IEEEproof}

\section{Applications}

In this Section, we use the derived closed-form expressions for the bivariate Nakagami-$m$ CDF to analyze some scenarios of interest in communications: (1) outage probability analysis of a dual-branch selection combining scheme in correlated Nakagami-$m$ fading, (2) calculation of higher order statistics of sampled Nakagami$-m$ fading channels.

\label{Applications}
\subsection{Outage probability of dual-branch SC}

Let us consider a dual-branch receiver affected by correlated Nakagami-$m$ fading. In this scenario, the expression for the instantaneous SNR per branch is given by $\gamma_i=E_s R_i^2/N_0$, where $i=1,2$, $E_s$ is the transmitted symbol energy and $N_0$ is the noise power spectral density. Equivalently, the average SNR per branch is given by $\bar{\gamma}_i= \text{E}\{R_i^2\}E_s/N_0$.

In a SC scheme, the receiver chooses the branch with maximum $\gamma_i$ for symbol decision i.e. $\gamma_{SC}=\text{max}\{\gamma_1,\gamma_2\}$. Hence, the OP of a dual-branch SC scheme affected by correlated Nakagami-$m$ fading can be expressed in terms of the bivariate CDF as
\begin{align}
P_{SC}(\gamma)&=\Pr\{\gamma_1<\gamma,\gamma_2<\gamma\}\nonumber \\&=F_{R_1,R_2}\left({r_1=\sqrt\gamma_1,r_2=\sqrt\gamma_2};m,\rho\right),
\end{align}
where $\rho=\rho_{R_1,R_2}$ is the power correlation coefficient according the definition in (\ref{powcorr}), $\Omega_1=\bar{\gamma_1}$ and $\Omega_2=\bar{\gamma_2}$ Substituting into (\ref{CDF01}), the following closed-form expression for the OP in this scenario is obtained

\begin{align}
\label{OP-fin}
P_{SC}(\gamma)&=1- \sum\limits_{k = 0}^{m - 1} \left[{\exp \left( { - \frac{{m\gamma }}{{\bar{\gamma} _1 }}} \right)\left[ {\left( {\frac{{m\gamma }}{{\bar{\gamma} _1 }}} \right)^k \frac{1}{{k!}} - 1} \right] + \exp \left( { - \frac{{m\gamma }}{{\bar{\gamma} _2 }}} \right)\left( {\frac{{m\gamma }}{{\bar{\gamma} _2 }}} \right)^k \frac{1}{{k!}}} \right. \nonumber\\
 &- \left( {1 - \rho } \right)^{ - k} \exp \left( { - \frac{m}{{1 - \rho }}\left( {\frac{\gamma}{\bar{\gamma}_1}  + \frac{\gamma}{\bar{\gamma}_2} } \right)} \right)  \times \nonumber \\ \nonumber
&\left\{ \left( \frac{m \gamma }{\bar{\gamma}_2}\right)^k\frac{1}{{\Gamma \left( {k + 1} \right)}}\Phi _3 \left( {1,k + 1;\tfrac{{\gamma }}{{\bar{\gamma}_2}}\tfrac{{m}}{{(1 - \rho )}},\rho \left( \tfrac{{\gamma}}{\sqrt{\bar{\gamma}_1 \bar{\gamma}_2}} \tfrac{{m}}{(1 - \rho) } \right)^2 } \right) \right. - \\ & \left.\left. \sum\limits_{i = 1}^{m - k} \left( \frac{m \gamma } {\bar{\gamma}_2}\right)^{k+i-1}\frac{1}{{\Gamma \left( {k + i} \right)}}\Phi _3 \left( {i,k + i;\tfrac{{\gamma }}{{\bar{\gamma}_2}}\tfrac{{m\rho}}{{(1 - \rho )}},\rho \left( \tfrac{{\gamma}}{\sqrt{\bar{\gamma}_1\bar{\gamma}_2} }\tfrac{{m}}{(1 - \rho) } \right)^2} \right)\right\} \right].
\end{align}

This closed-form expression for the OP of SC diversity in correlated non-identical Nakagami-$m$ fading is equivalent to those given the literature, either in terms of infinite series \cite{Tan1997} or in integral form \cite{Beaulieu2011}.

\subsection{Higher order statistics of sampled Nakagami-m fading channels}

Second-order statistics of fading channels incorporate information related with the dynamics of the random processes, characterizing their evolution due to the variation along a certain dimension (e.g. time). Specifically, the level crossing rate (LCR) measures how often the envelope fading crosses a threshold value, whereas the average fade duration (AFD) informs about the amount of time that the envelope remains below this threshold.

Classically, the LCR and the AFD have been calculated using Rice's approach \cite{Rice1944}, based upon the knowledge of the statistics of the continuous fading envelope and its time derivative. For the particular case of Nakagami-$m$ fading, the higher order statistics were calculated in \cite{Yacoub1999} in closed-form.

Recently \cite{Lopez2012}, it has been proposed an alternative formulation for the calculation of higher order statistics of sampled fading channels which incorporates the inherent discrete-time nature of fading channels due to sampling. Since the probability of missing a level crossing of the continuous envelope fading in a sampling interval
is not negligible for a finite sampling period \cite{Woods1982}, the LCR for a continuous fading process (i.e., using \cite{Rice1944}) can be seen as an upper bound of the associated sampled random process. Using this new approach \cite{Lopez2012}, we provide exact closed-form expressions for the LCR and the AFD of a sampled Nakagami-$m$ fading process.

The LCR of a sampled random process, defined as the average rate at which the envelope $R$ crosses a certain threshold $u$ in the positive (or equivalently in the negative) direction can be expressed as

\begin{equation}
\label{LCR_D} N_R(u)=\frac{\Pr\{R_1<u,R_2>u\}}{T}
\end{equation}
where $R_1\triangleq R(t)$, $R_2\triangleq R(t+T)$, and $T$ denotes the sampling period. $R_1$ and $R_2$ are correlated and identically distributed random variables, with a CDF denoted as $F_R(x)\triangleq F_{R_1}(x)=F_{R_2}(x)$, and joint CDF defined as $F_{R_1,R_2}(x,y)=\Pr\{R_1\leq x,R_2\leq y\}$.

In this scenario, the LCR of a sampled random process can be expressed in compact form as \cite[eq. 4]{Lopez2012}

\begin{equation}
\label{LCR}
N_R(u)=\frac{F_R(u)-F_{R_1,R_2}(u,u)}{T}=\frac{\bar{F}_R(u)-\bar{F}_{R_1,R_2}(u,u)}{T}.
\end{equation}

The AFD of a sampled random process, defined as the average duration of the envelope $R$ remaining below a specified threshold level $u$, can be calculated as \cite[eq. 6]{Lopez2012}

\begin{equation}
\label{AFD}
A_R(u)=\frac{\Pr\{R\leq u\}}{N_R(u)}=T\left({1-\frac{F_{R_1,R_2}(u,u)}{F_R(u)}}\right)^{-1}.
\end{equation}

For the particular case of Nakagami-$m$ fading, the exact closed-form expressions for the LCR and the AFD can be obtained setting $\{r_1/\sqrt{\Omega_1}=r_2/\sqrt{\Omega_2}=u\}$\footnote{where ${\Omega_1=\Omega_2=\Omega}$ and $u$ is the threshold value normalized to $\sqrt{\Omega}$.} in (\ref{CDF01}), and substituting into (\ref{LCR}) and (\ref{AFD}) respectively. These expressions for the LCR and AFD of sampled Nakagami-$m$ fading are new in the literature.

\section{Numerical Results}
\label{Numerical}

In this Section, we evaluate the expressions derived in the previous Section in some scenarios of interest. For the evaluation of the bivariate confluent hypergeometric function $\Phi_3$, we use the Mathematica$\texttrademark$ program given in Appendix \ref{App0}. The Nakagami-$m$ random variables used in Monte Carlo (MC) simulations have been generated considering that the correlation between the pairs of underlying Rayleigh processes composing the two Nakagami-$m$ signals occurs only between in-phase components (and equivalently between quadrature components) \cite{deSouza2008}; i.e., there is no cross-correlation between in-phase and quadrature components.

\begin{figure}[ht]
\begin{center}
\includegraphics[width=1.1\columnwidth]{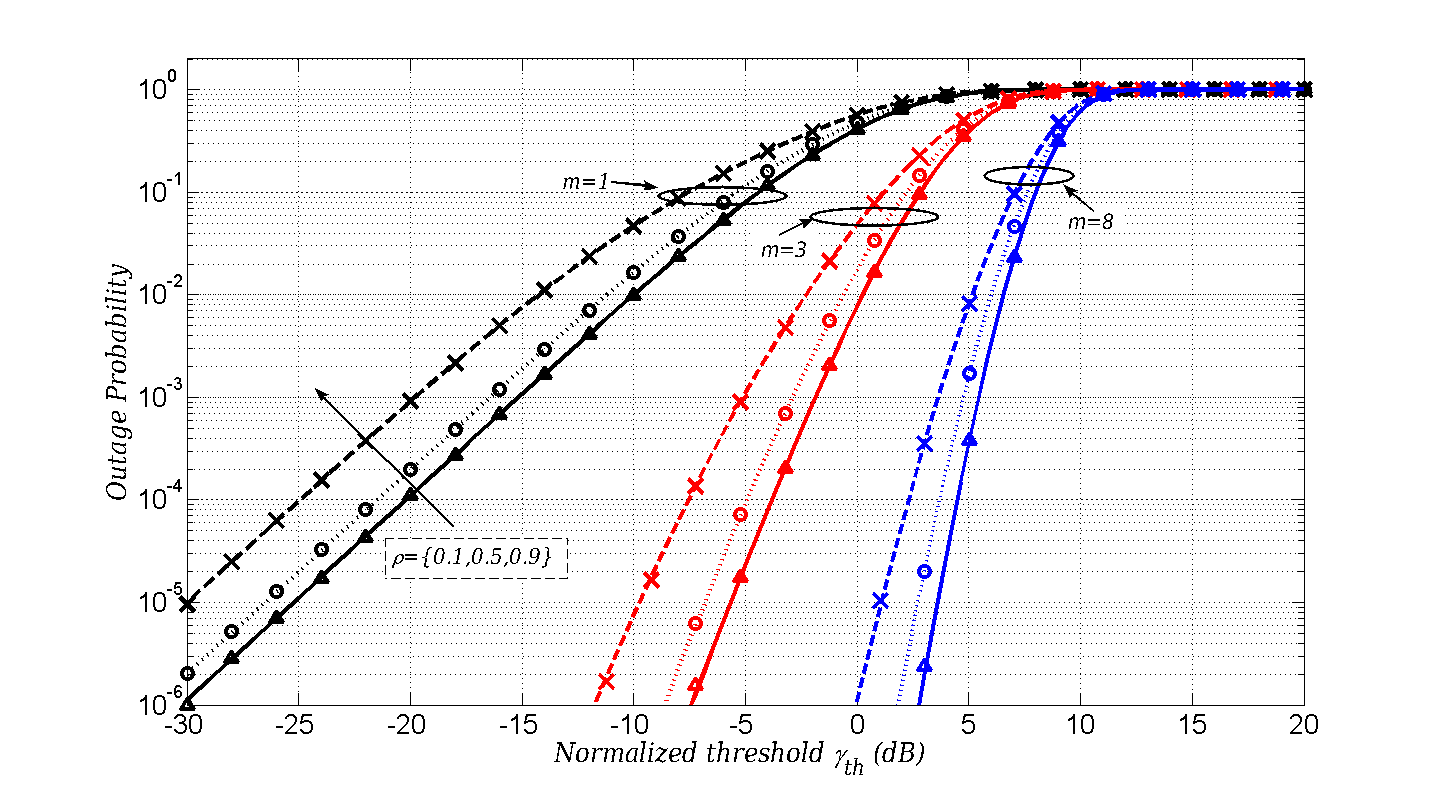}
\caption{Outage probability versus threshold level $\gamma_{th}$ (normalized to $\bar{\gamma}_1$), for balanced reception $\bar{\gamma}_1=\bar{\gamma}_2$ and different values of $m$ and $\rho$. Markers correspond to Monte Carlo simulations.}
\label{F3}
\end{center}
\end{figure}

Fig. \ref{F3} illustrates the OP of a dual-branch SC scheme with balanced branches (i.e.,$\bar{\gamma}_1$=$\bar{\gamma}_2$) affected by Nakagami-$m$ fading, for different values of fading severity index $m$ and power correlation coefficient $\rho$. The normalized SNR threshold $\gamma_{th}$ is defined as $\gamma_{th}\triangleq m\gamma/\bar{\gamma}_1$. It is appreciated how the closed-form expression and the MC simulations perfectly match. As expected, the OP is reduced as the fading severity index $m$ is increased, or the correlation between the reception branches is reduced.

\begin{figure}[ht]
\begin{center}
\includegraphics[width=1.1\columnwidth]{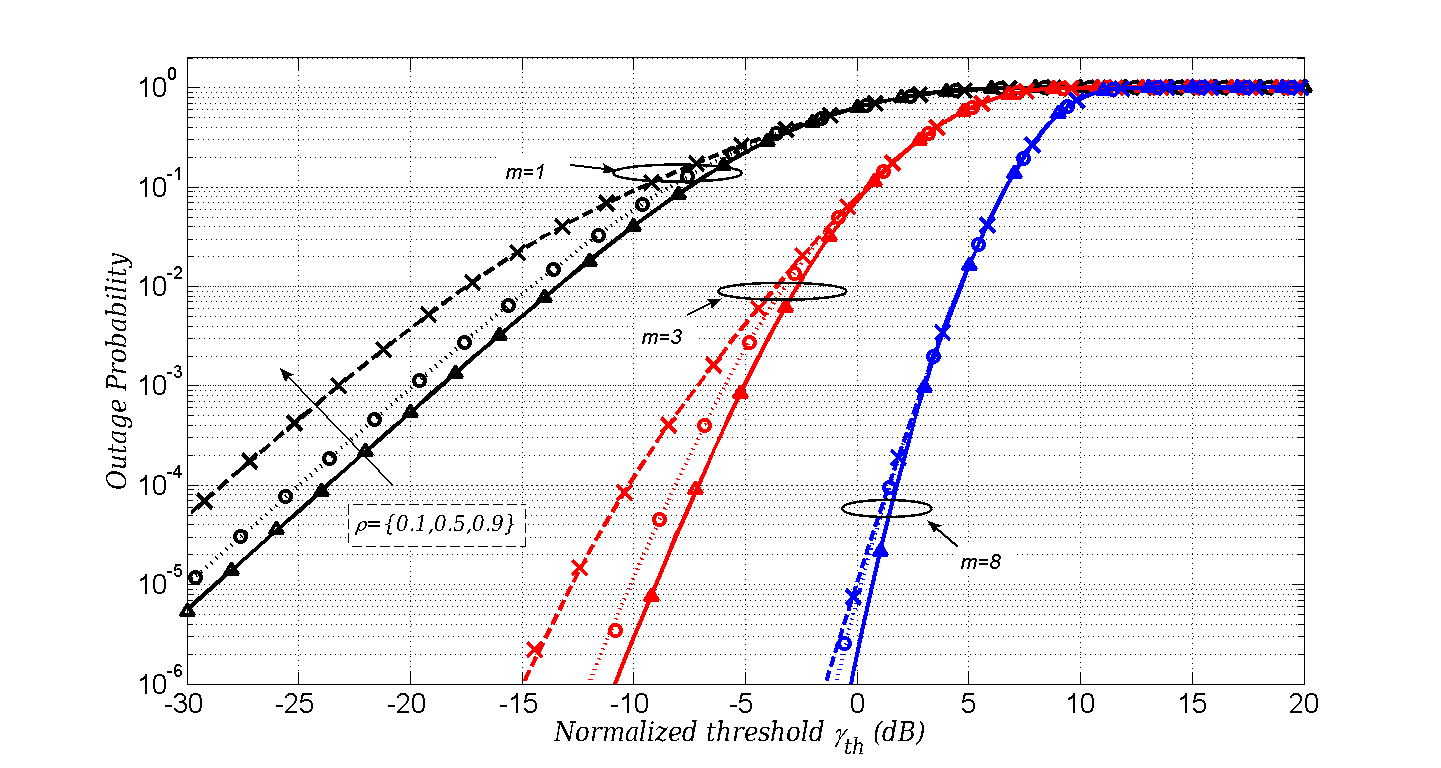}
\caption{Outage probability versus threshold level $\gamma_{th}$ (normalized to $\bar{\gamma}_1$), for unbalanced reception $\bar{\gamma}_1=5\bar{\gamma}_2$ and different values of $m$ and $\rho$. Markers correspond to Monte Carlo simulations.}
\label{F4}
\end{center}
\end{figure}

Fig. \ref{F4} shows the OP of a dual-branch SC scheme with unbalanced branches (i.e.,$\bar{\gamma}_1$=$5\bar{\gamma}_2$) affected by Nakagami-$m$ fading, for different values of fading severity index $m$ and power correlation coefficient $\rho$. The normalized SNR threshold is defined as $\gamma_{th}\triangleq m\gamma/\bar{\gamma}_1$. In this situation, the outage performance is lower compared to the scenario in Fig. \ref{F3}. Again, theoretical values and MC simulations are in excellent agreement.

\begin{figure}[ht]
\begin{center}
\includegraphics[width=1.1\columnwidth]{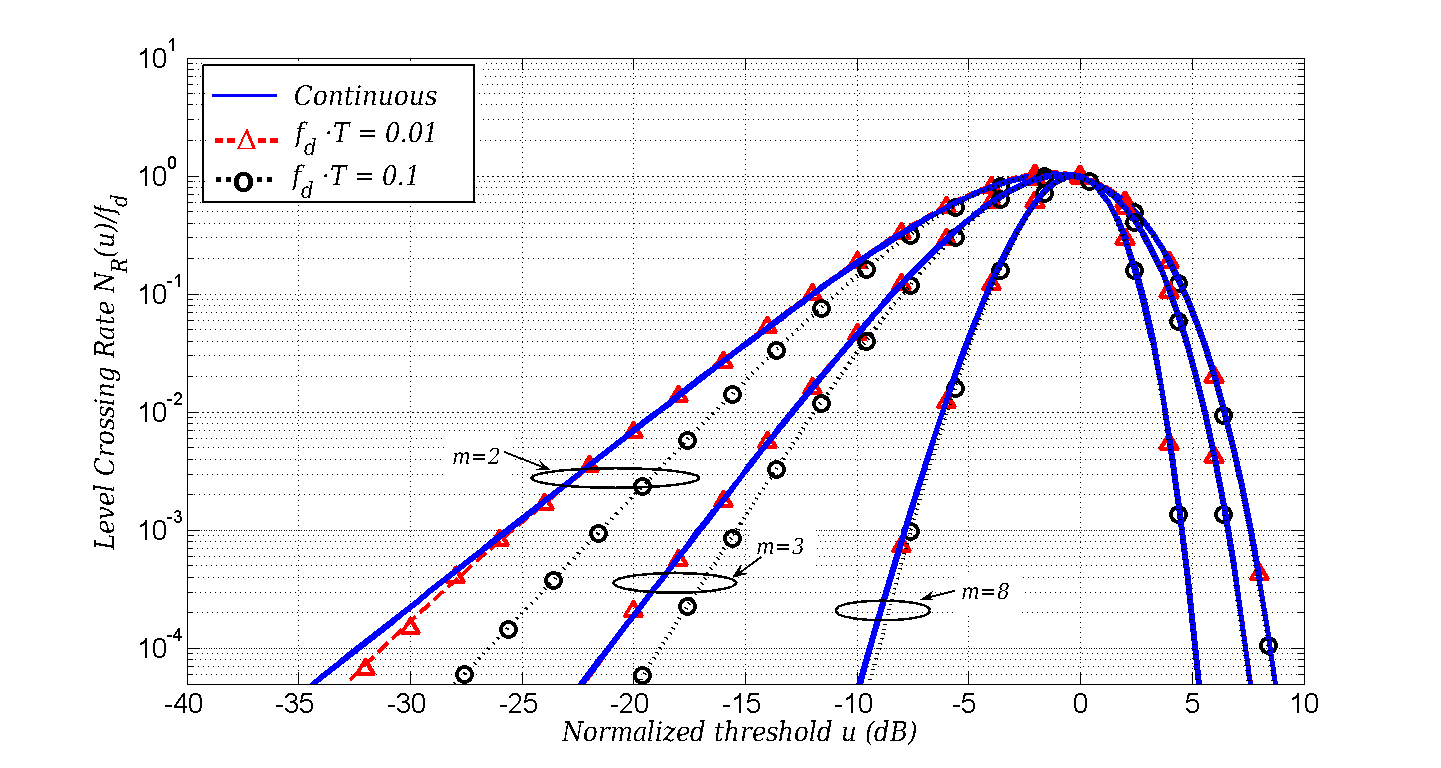}
\caption{Normalized level crossing rate $N_R(u)/f_d$ of a sampled Nakagami-$m$ fading channel, for different values of $f_d · T$ and $m$. Markers correspond to Monte Carlo simulations.}
\label{F5}
\end{center}
\end{figure}

In Fig. \ref{F5}, the level crossing rate of a sampled Nakagami-$m$ fading process is depicted, for different values of Doppler frequency $f_d$ and sampling period $T$. We have assumed that the underlying Rayleigh random variables have a correlation coefficient {$\rho_{R}(\tau)=J_0(2\pi f_d \tau)$}, which yields an envelope fading normalized correlation coefficient \cite{Iskander2005} given by

\begin{equation}
\rho=\frac{\tfrac{1}{m}\left[\left(\tfrac{\Gamma(m+1/2)}{\Gamma(m)}\right)^2 {}_{2}F_1\left(-0.5,-0.5,m,\rho_R\right)-1\right]}{1-\tfrac{1}{m}\left(\tfrac{\Gamma(m+1/2)}{\Gamma(m)}\right)^2},
\end{equation}
where ${}_2 F_1$ is the Gauss hypergeometric function. It is observed that the LCR values of the sampled Nakagami$-m$ envelope differ from the LCR values of the continuous Nakagami-$m$ envelope \cite{Yacoub1999} for low values of the threshold level. This difference is increased as the value of $f_d · T$ product is increased, i.e., as the fading process is less oversampled. Interestingly, the continuous LCR is an upper bound of the actual LCR of the sampled fading process, as indicated in \cite{Lopez2012}.

\begin{figure}[ht]
\begin{center}
\includegraphics[width=0.6\columnwidth]{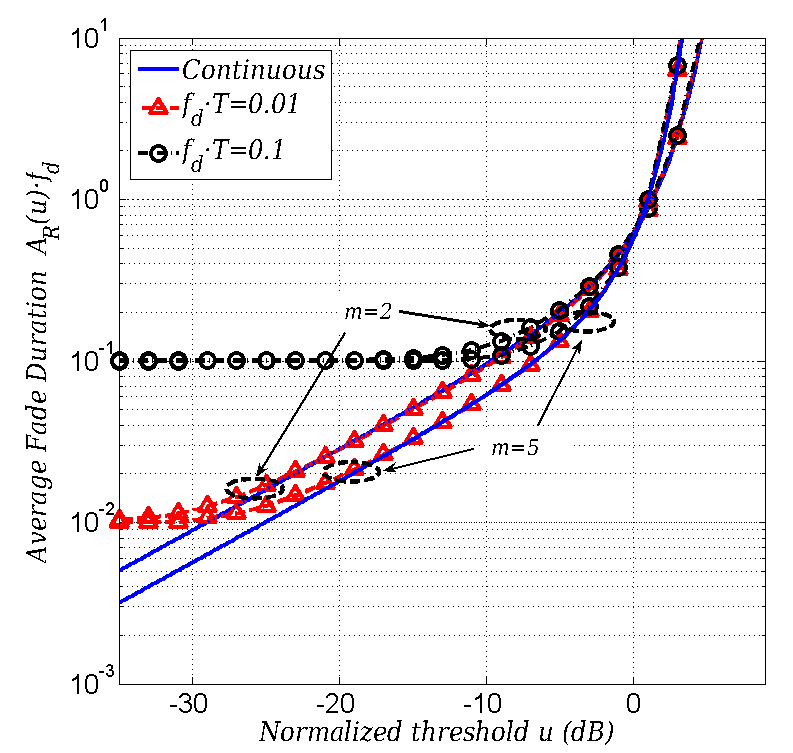}
\caption{Normalized average fade duration $A_R(u)·f_d$ of a sampled Nakagami-$m$ fading channel, for different values of $f_d · T$ and $m$.}
\label{F6}
\end{center}
\end{figure}

Finally, Fig. \ref{F6}, illustrates the average fade duration of a sampled Nakagami-$m$ fading process, for the same values of Doppler frequency $f_d$ and sampling period $T$ used in Fig. \ref{F5}. In this case, the AFD of the continuous process is a lower bound of the actual AFD. It is observed that the AFD of the sampled Nakagami-$m$ process reaches an irreducible floor equal to $T$; this is in concordance with the definition of the AFD for a sampled fading process, as the minimum expectable duration of a fading is necessarily one sampling period interval.

\section{Conclusion}
\label{Conclusion}
In this paper, we have derived exact closed-form expressions for the bivariate Nakagami-$m$ CDF with positive integer $m$, in terms of the confluent hypergeometric function of two variables $\Phi_3$. These expressions are particularly convenient, since $\Phi_3$ function is well studied in classical books of integrals, special functions and Laplace transforms, and can be efficiently computed using a numerical Laplace inversion method. Besides, the derived CDF is shown to reduce to the bivariate Rayleigh CDF given in terms of the Marcum-$Q$ function, which suggests a further connection between this family of hypergeometric functions and the generalized $Q_m$ functions.

We have illustrated the applicability of the obtained expression in two scenarios of interest in communications: an outage probability analysis of a dual-branch SC in correlated Nakagami-$m$ fading channels, and the calculation of the higher order statistics of sampled Nakagami-$m$ fading channels. Simulations corroborate the validity of the derived closed-form expressions.

\section*{Acknowledgements}
\label{Ack}

This work was supported in part by Junta de Andalucia (project ``Analysis and design of cooperative adaptive MIMO-OFDM systems''), Spanish Government-FEDER (TEC2010-18451, TEC2011-25473), the University of Malaga and European Union under Marie-Curie COFUND U-mobility program (ref. 246550), and by the company AT4 Wireless. The work of Dr. Morales-Jimenez is supported by Spanish Ministry of Science and Innovation (CSD2008-00010, COMONSENS).

\appendices

\section{Proof of proposition \ref{prop3}}
\label{App3}

We aim to find a closed-form expression for the integral $\H_m \left( {\gamma ,\delta ,u} \right)$, defined as
\begin{equation}
\H_m\left(u,\gamma,\delta\right) \triangleq {\int_u^\infty  {x^{2m - 1} \exp \left( { - m x^2 } \right)Q_m \left( {\gamma x,\delta } \right)dx}}.
\end{equation}

The Marcum-$Q$ function can be expressed in terms of a contour integral \cite{Proakis2001} as
\begin{equation}
Q_m \left( {\gamma x,\delta } \right) = \exp \left( { - \frac{{\gamma ^2 x^2  + \delta ^2 }}{2}} \right)\oint_\Gamma  {\frac{1}{{p^m }}\frac{1}{{1 - p}}\exp \left\{ {\frac{{\gamma ^2 x^2 }}{{2p}} + \frac{{\delta ^2 p}}{2}} \right\}dp} ;
\end{equation}
where $\Gamma$ is a circular contour of radius less than unity enclosing the origin. Thus, we can express
\begin{equation}
\H_m\left(u,\gamma,\delta\right) = \int_u^\infty  {x^{2m - 1} \exp \left( { - m x^2 } \right)\left\{ {\exp \left( { - \frac{{\gamma ^2 x^2  + \delta ^2 }}{2}} \right)\oint_\Gamma  {\frac{1}{{p^m }}\frac{1}{{1 - p}}\exp \left\{ {\frac{{\gamma ^2 x^2 }}{{2p}} + \frac{{\delta ^2 p}}{2}} \right\}dp} } \right\}dx} .
\end{equation}

After some manipulations, and using $\alpha = \gamma ^2  + 2m$, we have
\begin{equation}
\label{eq1app3}
\H_m\left(u,\gamma,\delta\right) = \exp \left( { - \frac{{\delta ^2 }}{2}} \right)\int_u^\infty  {x^{2m - 1} \exp \left( { - \frac{{x^2 }}{2}\alpha } \right)\left\{ {\oint_\Gamma  {\frac{1}{{p^m }}\frac{1}{{1 - p}}\exp \left\{ {\frac{{\gamma ^2 x^2 }}{{2p}} + \frac{{\delta ^2 p}}{2}} \right\}} dp} \right\}dx} .
\end{equation}

Changing the integration order in (\ref{eq1app3}), we obtain
\begin{equation}
\label{eq2app3}
\H_m\left(u,\gamma,\delta\right)= \exp \left( { - \frac{{\delta ^2 }}{2}} \right)\oint_\Gamma  {\left\{ {\int_u^\infty  {x^{2m - 1} \exp \left( { - \frac{{x^2 }}{2}\varepsilon } \right)dx} } \right\}\frac{1}{{p^m }}\frac{1}{{1 - p}}\exp \left( {\frac{{\delta ^2 p}}{2}} \right)dp;}
\end{equation}

where $\varepsilon  \buildrel \Delta \over = \alpha  - \frac{{\gamma ^2 }}{p}$

The inner integral in (\ref{eq2app3}) can be expressed as \cite{Kapinas2009}
\begin{equation}
\label{aa1}
\int_u^\infty  {x^{2m - 1} \exp \left( { - \frac{{x^2 }}{2}\varepsilon } \right)dx}  = \varepsilon ^{ - m} 2^{m - 1} \Gamma \left( {m,\frac{{u^2 \varepsilon }}{2}} \right),
\end{equation}
where $\Gamma(m,w)$ is the upper incomplete Gamma function. In case $m$ is a positive integer, $\Gamma(m,w)$ can be expressed as 

\begin{equation}
\label{aa2}
\Gamma \left( {m,w} \right) = \left( {m - 1} \right)!\exp \left( { - w} \right)\sum\limits_{k = 0}^{m - 1} {\frac{{w^k }}{{k!}}} ;
\end{equation}

Substituting (\ref{aa1}) and (\ref{aa2}) into (\ref{eq2app3}) we have
\begin{align}
\H_m\left(u,\gamma,\delta\right) &= \exp \left( { - \frac{{\delta ^2 }}{2}}\right) \alpha ^{ - m}2^{m - 1} \left( {m - 1} \right)! \oint_\Gamma   \frac{1}{{p^m }}\frac{1}{{1 - p}}\exp \left( {\frac{{\delta ^2 p}}{2}} \right) \times \nonumber\\ &{\left\{ { \left( {1 - \frac{{\gamma ^2 }}{{\alpha p}}} \right)^{ - m} \exp \left( { - \frac{{\alpha u^2 }}{2}\left( {1 - \frac{{\gamma ^2 }}{{\alpha p}}} \right)} \right)\sum\limits_{k = 0}^{m - 1} {\left( {\frac{{\alpha u^2 }}{2}} \right)^k \left( {1 - \frac{{\gamma ^2 }}{{\alpha p}}} \right)^k \frac{1}{{k!}}} } \right\}}dp,
\end{align}
which can be conveniently rearranged as 
\begin{equation}
\H_m\left(u,\gamma,\delta\right) = \exp \left( { - \frac{{\delta ^2 }}{2}} \right)\exp \left( { - \alpha \frac{{u^2 }}{2}} \right)2^{m - 1} \alpha ^{ - m} \left( {m - 1} \right)!\sum\limits_{k = 0}^{m - 1} {\left( {\alpha \frac{{u^2 }}{2}} \right)^k \frac{1}{{k!}}} I_{m,k} \left( {u,\delta ,\gamma } \right)
\end{equation}
where
\begin{equation}
I_{m,k}  =  - \oint_\Gamma  {\frac{1}{{\left( {p - \rho } \right)^{(m - k)} }}\frac{1}{{\left( {p - 1} \right)}}p^{ - k} \exp \left( {\frac{{u^2 }}{2}\frac{{\gamma ^2 }}{p}} \right)\exp \left( {\frac{{\delta ^2 p}}{2}} \right)dp}
\end{equation}
and $\rho=\tfrac{\gamma^2}{\gamma^2+2m}$.

Using partial fraction expansion, it is possible to re-express $I_{m,k}$ as follows, in order to separate the presence of the singularities at $p = 1$ and $p = \rho$ into two different integrals. Thus
\begin{equation}
\label{eqExpandida}
I_{m,k}  = \oint_\Gamma  {\frac{{A_0 }}{{p - 1}}p^{ - k} \exp \left( {\frac{{u^2 }}{2}\frac{{\gamma ^2 }}{p}} \right)\exp \left( {\frac{{\delta ^2 p}}{2}} \right)dp}  + \sum\limits_{i = 1}^{m - k} {\oint_\Gamma  {\frac{{A_i }}{{\left( {p - \rho } \right)^i }}p^{ - k} \exp \left( {\frac{{u^2 }}{2}\frac{{\gamma ^2 }}{p}} \right)\exp \left( {\frac{{\delta ^2 p}}{2}} \right)dp}}
\end{equation}
where
\begin{equation}
A_0  =  - \left( {\frac{1}{{1 - \rho }}} \right)^{(m - k)} ,A_i  = \left( {\frac{1}{{1 - \rho }}} \right)^{m-k-i+1}.
\end{equation}

Finally, identifying the integrals (\ref{def2}) and (\ref{def3}) in (\ref{eqExpandida}), yields the desired expression (\ref{eqprop3}).

\section{Proof of proposition \ref{prop1}}
\label{App1}

Let us express the integral $\mathcal{I}_k\left(u,\gamma,\delta\right)$ in the following compact form as
 
\begin{equation}
\mathcal{I}_k\left(u,\gamma,\delta\right)= \oint_\Gamma F_1(p) \exp \left( {\frac{{\delta ^2 p}}{2}} \right)dp,
\end{equation}
where the contour $\Gamma$ is defined as a circular contour of radius less than unity that encloses the origin and

\begin{equation}
F_1(p)\triangleq\frac{1}{{p - 1}}p^{ - k} \exp \left( {\frac{{u^2 }}{2}\frac{{\gamma ^2 }}{p}} \right).
\end{equation}

\begin{figure}[ht]
\begin{center}
\includegraphics[width=.89\columnwidth]{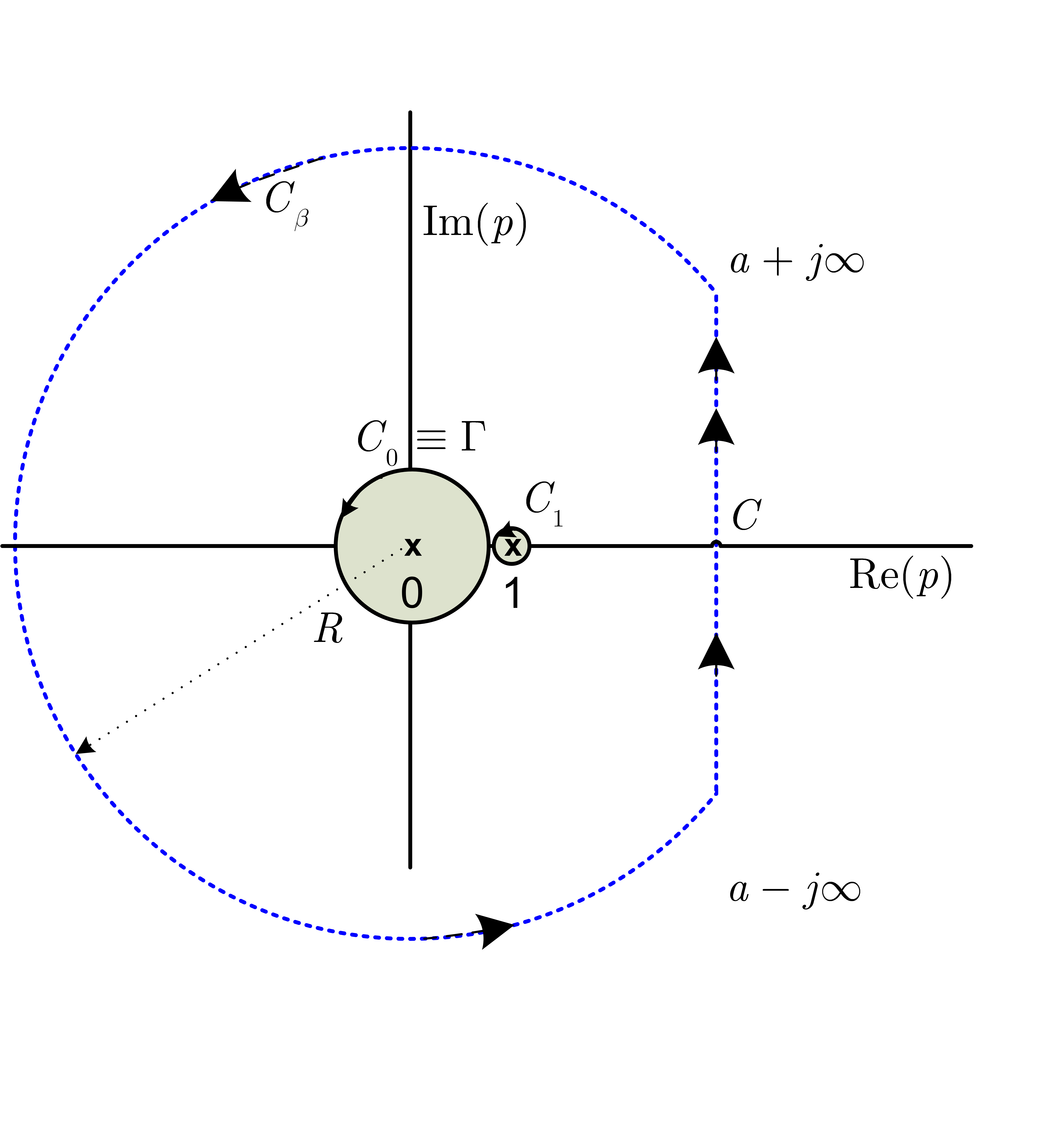}
\caption{Contour integration for integral $\mathcal{I}_{k}$.}
\label{F1}
\end{center}
\end{figure}

Interestingly, the integrand in $\mathcal{I}_k$ is in the form of an inverse Laplace transform. Hence, we aim to find a connection between the integral defined in the contour $\Gamma$ and the general Bromwich integral given by

\begin{equation}
\label{Bromwich}
f(g) = \frac{1}{{2\pi j}}\int_{a - j\infty }^{a + j\infty } {F(p)\exp \left( {gp} \right)dp}=\mathcal{L}^{-1}\left\{F(p)\right\},
\end{equation}
where $F(p)$ is the Laplace transform of $f(g)$. Let us consider the contour $C$ given in Fig. \ref{F1}, where the value of $a$ is chosen to be at the right of all the singularities of $F_1(p)$. Hence, the contour integral along $C$ is given by

\begin{equation}
\label{eqappx}
\oint_C {F_1 (p)\exp \left( {\frac{{\delta ^2 p}}{2}} \right)dp}  = \frac{1}{{2\pi j}}\int_{a - j\infty }^{a + j\infty } {F_1 (p)\exp \left( {\frac{{\delta ^2 p}}{2}} \right)dp}  + \oint_{C_\beta  } {F_1 (p)\exp \left( {\frac{{\delta ^2 p}}{2}} \right)dp}.
\end{equation}

Using Cauchy-Goursat theorem, we can equivalently express

\begin{equation}
\label{eqapp2}
\oint_C {F_1 (p)\exp \left( {\frac{{\delta ^2 p}}{2}} \right)dp}  = \oint_{C_0 } {F_1 (p)\exp \left( {\frac{{\delta ^2 p}}{2}} \right)dp}  + \oint_{C_1 } {F_1 (p)\exp \left( {\frac{{\delta ^2 p}}{2}} \right)dp}
\end{equation}
where $C_0$ and $C_1$ are closed contours which enclose the singularities at $p=0$ and $p=1$ respectively. Combining (\ref{eqappx}) and (\ref{eqapp2}), we can express

\begin{equation}
\frac{1}{{2\pi j}}\int_{a - j\infty }^{a + j\infty } {F_1 (p)\exp \left( {\frac{{\delta ^2 p}}{2}} \right)dp}  + \oint_{C_\beta  } {F_1 (p)\exp \left( {\frac{{\delta ^2 p}}{2}} \right)dp}  = \oint_{C_0 + C_1 } {F_1 (p)\exp \left( {\frac{{\delta ^2 p}}{2}} \right)dp}.
\end{equation}

The modulus of $F_1(p)$ in $C_{\beta}$ is given by 
\begin{equation}
\label{david01}
\left| {F_1 (p)} \right|_{p = {\mathop{\rm Re}\nolimits} ^{j\theta } } = \frac{1}{{\left| {p - 1} \right|}}\frac{1}{{\left| p \right|^k }}\left| {e^{b/p} } \right|,
\end{equation}
with $b=u^2\gamma^2/2$. Using
\begin{equation}
\frac{1}{{\left| {p - 1} \right|}} \le \frac{1}{{\left| {\left| p \right| - 1} \right|}}\mathop  \le \limits_{R > 2} \frac{2}{R},
\end{equation}
and considering that
\begin{equation}
 \left| {e^{b/p} } \right| = \left| {e^{{\mathop{\rm Re}\nolimits} \left( {b/p} \right)} } \right| \le e^{\left| {b/R} \right|} \mathop  \le \limits_{R > R_0 } e^{\left| {b/R_0 } \right|},
\end{equation}
we have
\begin{equation}
\left| {F_1 (p)} \right|\mathop  \le \limits_{R > R_0 } \underbrace {2e^{\left| {b/R_0 } \right|} }_M\underbrace {R^{ - \left( {k + 1} \right)} }_{R^{ - l} }.
\end{equation}

Thus, $\left| {F_1 (p)} \right|_{p=Re^{j\theta}} \le MR^{ - l}$ for some $l>0$ on $C_{\beta}$ as $R \to \infty$, and using Jordan's lemma the integral $\oint_{C_\beta}$ equals to 0. Hence, using the residue theorem and choosing $C_0  \equiv \Gamma$, we have


\begin{align}
\label{eqrel1}
\mathcal{I}_k\left(u,\gamma,\delta\right) &= \frac{1}{{2\pi j}}\int_{a - j\infty }^{a + j\infty } {F_1 (p)\exp \left( {\frac{{\delta ^2 p}}{2}} \right)dp}  - {\mathop{\rm Re}\nolimits} s\left\{ {F_1 (p)\exp \left( {\frac{{\delta ^2 p}}{2}} \right)} \right\}_{p = 1}\\
  &= {\cal L}^{ - 1} \left\{ {F_1 (p)} \right\}_{g = \frac{{\delta ^2 }}{2}}  - {\text{Res}}\left\{ {F_1 (p)\exp \left( {\frac{{\delta ^2 p}}{2}} \right)} \right\}_{p = 1},\nonumber
\end{align}
where $\text{Res}\left\{F(p)\right\}_{p=x}$ denotes the residue of $F(p)$ at $p=x$.

Expression (\ref{eqrel1}) defines a relationship between the integral $\mathcal{I}_k$ defined in a closed contour $\Gamma$ and the inverse Laplace transform of the integrand. Using \cite[4.24.9]{Erdelyi1954}, we have

\begin{equation}
\label{relLaplace}
{\cal L}^{ - 1} \left\{ {F_1 (p)} \right\}_{g = \frac{{\delta ^2 }}{2}}  = \frac{{1 }}{{\Gamma (k + 1)}}\left( {\frac{{\delta ^2 }}{2}} \right)^k \Phi _3 \left( {1,k + 1;\frac{{\delta ^2 }}{2},\frac{{u^2 }}{2}\frac{{\gamma ^2 \delta ^2 }}{2}} \right)
\end{equation}

where $\Phi _3 \left( {\alpha ,\beta ;x,y} \right)$ is the confluent hypergeometric function of two variables \cite[9.261.3]{Gradstein2007}, which is one of the bivariate forms of the confluent hypergeometric function $_1F_1$. Using (\ref{relLaplace}), we can express the integral $\mathcal{I}_k$ as

\begin{equation}
\label{app1final}
\mathcal{I}_k\left(u,\gamma,\delta\right)  = \frac{{1 }}{{\Gamma (k + 1)}}\left( {\frac{{\delta ^2 }}{2}} \right)^k \Phi _3 \left( {1,k + 1;\frac{{\delta ^2 }}{2},\frac{{u^2 }}{2}\frac{{\gamma ^2 \delta ^2 }}{2}} \right) - \exp \left( {\frac{{\gamma ^2 u^2 }}{2} + \frac{{\delta ^2 }}{2}} \right).
\end{equation}

\section{Proof of proposition \ref{prop2}}
\label{App2}

The calculation of $\mathcal{J}_{i,k}$ is performed using a similar procedure to that in Appendix \ref{App1}. Let us express the integral $\mathcal{J}_{i,k}\left(u,\gamma,\delta,\rho\right)$ in the following compact form as

\begin{equation}
\mathcal{J}_{i,k}\left(u,\gamma,\delta,\rho\right)= \oint_\Gamma F_{2,i}(p) \exp \left( {\frac{{\delta ^2 p}}{2}} \right)dp,
\end{equation}

where the contour $\Gamma$ is defined as a circular contour of radius less than unity that encloses the origin and

\begin{equation}
F_{2,i}(p)\triangleq\frac{1}{{(p - \rho)^i}}p^{ - k} \exp \left( {\frac{{u^2 }}{2}\frac{{\gamma ^2 }}{p}} \right).
\end{equation}

Again, the integrand in $\mathcal{J}_{i,k}$ is in the form of an inverse Laplace transform. Thus, we aim to find a connection between the integral defined in the contour $\Gamma$ and the general Bromwich integral (\ref{Bromwich}).

\begin{figure}[ht]
\begin{center}
\includegraphics[width=.89\columnwidth]{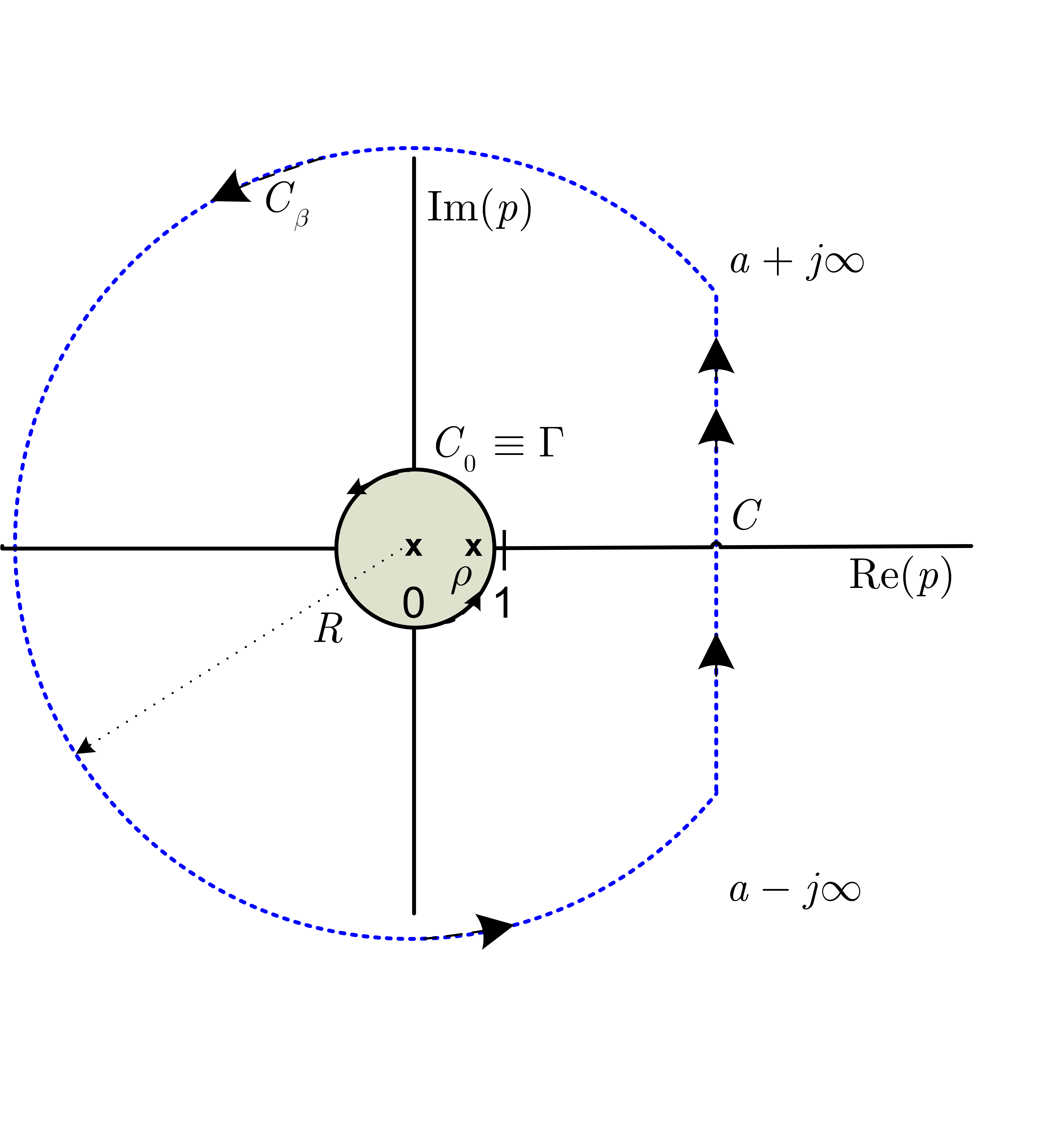}
\caption{Contour integration for integral $\mathcal{J}_{i,k}$.}
\label{F2}
\end{center}
\end{figure}

Let us consider the contour $C$ given in Fig. \ref{F2}, where the value of $a$ is chosen to be at the right of all the singularities of $F_{2,i}(p)$. Hence, the contour integral along $C$ is given by

\begin{equation}
\label{eqapp1}
\oint_C {F_{2,i} (p)\exp \left( {\frac{{\delta ^2 p}}{2}} \right)dp}  = \frac{1}{{2\pi j}}\int_{a - j\infty }^{a + j\infty } {F_{2,i} (p)\exp \left( {\frac{{\delta ^2 p}}{2}} \right)dp}  + \oint_{C_\beta  } {F_{2,i} (p)\exp \left( {\frac{{\delta ^2 p}}{2}} \right)dp}.
\end{equation}

As in (\ref{david01}), $\left| {F_{2,i} (p)} \right|_{p=Re^{j\theta}} \le MR^{ - l} $ for some $l>0$ on $C_{\beta}$ as $R \to \infty$, so that the integral $\oint_{C_\beta}$ equals to 0.

Under Cauchy - Goursat theorem, we can choose a contour $\Gamma$ as a circular contour of radius less than unity enclosing the origin. Since $\rho  < 1$, the singularities of $F_{2,i} (p)$ at $p = 0$ and $p = \rho$  are enclosed within $\Gamma$. Thus, we have

\begin{equation}
\frac{1}{{2\pi j}}\int_{a - j\infty }^{a + j\infty } {F_{2,i} (p)\exp \left( {\frac{{\delta ^2 p}}{2}} \right)dp}  = \oint_\Gamma  {F_{2,i} (p)\exp \left( {\frac{{\delta ^2 p}}{2}} \right)dp}
\end{equation}

Therefore, the integral $\mathcal{J}_{i,k}$ can be expressed as

\begin{equation}
\label{eqrel2}
\mathcal{J}_{i,k}  = {\cal L}^{ - 1} \left\{ {F_{2,i} (p)} \right\}_{g = \frac{{\delta ^2 }}{2}} ;
\end{equation}

The inverse Laplace transform in (\ref{eqrel2}) can be calculated using \cite[4.24.9]{Erdelyi1954}, yielding

\begin{equation}
\mathcal{J}_{i,k}\left(u,\gamma,\delta,\rho\right)= {\cal L}^{ - 1} \left\{ {F_{2,i} (p)} \right\}_{g = \frac{{\delta ^2 }}{2}}  = \frac{{1 }}{{\Gamma (k + i)}}\left( {\frac{{\delta ^2 }}{2}} \right)^{k + i - 1} \Phi _3 \left( {i,k + i;\rho \frac{{\delta ^2 }}{2},\frac{{u^2 }}{2}\frac{{\gamma ^2 \delta ^2 }}{2}} \right).
\end{equation}

\section{Proof of lemma \ref{lemma1}}
\label{App4}

The joint PDF of two correlated Nakagami$-m$ variates is given in \cite{Nakagami1960,Simon2005} by
\begin{align}
\label{eq1}
f_{R_1 ,R_2 } (r_1  ,r_2  |m,\rho ) &= \frac{{4m^{m + 1} \left( {r_1 r_2 } \right)^m }}{{\Gamma (m)\Omega _1 \Omega _2 \left( {1 - \rho } \right)\left( {\sqrt {\Omega _1 \Omega _2 \rho } } \right)^{m - 1} }}\exp \left( { - \frac{m}{{1 - \rho }}\left\{ {\frac{{r_1^2 }}{{\Omega _1 }} + \frac{{r_2^2 }}{{\Omega _2 }}} \right\}} \right)\times\\
&I_{m - 1} \left( {\frac{{2m\sqrt \rho  r_1 r_2 }}{{\sqrt {\Omega _1 \Omega _2 } \left( {1 - \rho } \right)}}} \right);
\nonumber
\end{align}
where $\Omega _i  = E\left\{ {r_i^2 } \right\}$, $I_n(x)$ is the $n^{th}$ order modified Bessel function of the first kind, and $\rho$ is the power correlation coefficient between $R_1$ and $R_2$ according to the definition in (\ref{powcorr}).

For the sake of compactness, we define
\begin{equation}
{{x = }}\frac{{r_1 }}{{\sqrt {\Omega _1 } }},{{y = }}\frac{{r_2 }}{{\sqrt {\Omega _2 } }};\kappa = \frac{{2m^m }}{{\Gamma (m)\left( {\sqrt \rho  } \right)^{m - 1}}};\alpha  = \frac{{2m}}{{\left( {1 - \rho } \right)}};
\end{equation}
and hence this PDF can be rewritten as
\begin{equation}
\label{eqPDF}
f_{X,Y} (x,y;m,\rho) = \kappa\alpha \left( {xy} \right)^m \exp \left( { - \frac{\alpha }{2}\left\{ {x^2  + y^2 } \right\}} \right)I_{m - 1} \left( {\alpha \sqrt \rho  xy} \right);
\end{equation}

The bivariate CDF is defined as

\begin{equation}
F_{X,Y} (u,v;m,\rho) = \Pr \left\{ {x \le u,y \le v} \right\} = \int_0^u {\int_0^v {f_{X,Y} (x,y)dydx} }
\end{equation}
or equivalently, the complementary CDF (CCDF)
\begin{equation}
\label{CCDF01}
\bar F_{X,Y} (u,v;m,\rho) = \Pr \left\{ {x > u,y > v} \right\} = \int_u^\infty  {\int_v^\infty  {f_{X,Y} (x,y)dydx} }
\end{equation}

Substituting (\ref{eqPDF}) into (\ref{CCDF01}), we have
\begin{align}
\label{CCDF02}
\bar F_{X,Y} (u,v;m,\rho) &= \kappa\alpha \int_u^\infty  {x^m \exp \left( { - \frac{\alpha }{2}x^2 } \right)\left\{ {\int_v^\infty  {y^m \exp \left( { - \frac{\alpha }{2}y^2 } \right)I_{m - 1} \left( {\alpha \sqrt \rho  xy} \right)dy} } \right\}dx} \\& = \kappa\alpha \int_u^\infty  {x^m \exp \left( { - \frac{\alpha }{2}x^2 } \right)\Upsilon \left( {x,v} \right)dx}
\nonumber
\end{align}
where
\begin{equation}
\Upsilon \left( {x,v} \right) = \int_v^\infty  {y^m \exp \left( { - \frac{\alpha }{2}y^2 } \right)I_{m - 1} \left( {\alpha \sqrt \rho  xy} \right)dy}.
\end{equation}

Let us first find a closed-form expression for $\Upsilon \left( {x,v} \right)$. Performing a change of variable 
\begin{equation}
\left[\kern-0.15em\left[ \begin{array}{l}
 \alpha y^2  = z^2  \\ 
 \sqrt \alpha  dy = dz \\ 
 \end{array} 
 \right]\kern-0.15em\right] \Rightarrow \Upsilon \left( {x,v} \right) = \int_{v\sqrt \alpha  }^\infty  {\left( {\frac{z}{{\sqrt \alpha  }}} \right)^m \frac{1}{{\sqrt \alpha  }}\exp \left( { - \frac{{z^2 }}{2}} \right)I_{m - 1} \left( {tz} \right)dz}.
\end{equation}
where $ {t= \sqrt {\alpha \rho } x }$. After some straightforward manipulations, we can express
\begin{equation}
\label{ups1}
\Upsilon \left( {x,v} \right) = \left( {\sqrt \alpha  } \right)^{ - m - 1} t^{m - 1} \exp \left( {\frac{{t^2 }}{2}} \right)\underbrace {\frac{1}{{t^{m - 1} }}\int_{v\sqrt \alpha  }^\infty  {z^m \exp \left( { - \frac{{t^2 }}{2}} \right)\exp \left( { - \frac{{z^2 }}{2}} \right)I_{m - 1} \left( {tz} \right)dz} }_{Q_m \left( {t,v\sqrt \alpha  } \right)}.
\end{equation}

Thus, the first integral can be expressed in terms of the Marcum-$Q$ function $Q_m \left( {a,b} \right)$ \cite{Simon2005}, i.e.
\begin{equation}
\label{ups2}
{\Upsilon \left( {x,v} \right) = \frac{1}{\alpha }\left( {\sqrt \rho  x} \right)^{m - 1} \exp \left( {\frac{{\alpha \rho x^2 }}{2}} \right)Q_m \left( {x\sqrt {\alpha \rho } ,v\sqrt \alpha  } \right)}.
\end{equation}

Substituting (\ref{ups2}) into (\ref{CCDF02}), we obtain
\begin{equation}
\bar F_{X,Y} (u,v;m,\rho) = \kappa\int_u^\infty  {x^m \exp \left( { - \frac{\alpha }{2}x^2 } \right)\left( {\sqrt \rho  x} \right)^{m - 1} \exp \left( {\frac{{\alpha \rho x^2 }}{2}} \right)Q_m \left( {x\sqrt {\alpha \rho } ,v\sqrt \alpha  } \right)dx}
\end{equation}
which can be reexpressed in compact form as
\begin{equation}
\label{CCDF03}
\bar F_{X,Y} (u,v;m,\rho) = \kappa\left( {\sqrt \rho  } \right)^{m - 1} \int_u^\infty  {x^{2m - 1} \exp \left( { - m x^2 } \right)Q_m \left( {\gamma x,\delta } \right)dx},
\end{equation}
where
\begin{align}
\gamma  = \sqrt {\alpha \rho }, \\
\delta  = v\sqrt \alpha.
\end{align}

The integral in (\ref{CCDF03}) can be identified with (\ref{eqdef1}), yielding
\begin{equation}
\label{CCDF04}
\bar F_{X,Y} (u,v;m,\rho) = \kappa\left( {\sqrt \rho  } \right)^{m - 1} \H_m\left(u,\gamma,\delta\right).
\end{equation}

Finally, substituting $\bar{F}_{R_1 ,R_2 } \left( {r_1,r_2;m,\rho } \right) =\bar{F}_{X ,Y } \left( {u=r_1/\sqrt{\Omega_1},v=r_2/\sqrt{\Omega_2};m,\rho } \right)$ the desired expression for the bivariate Nakagami$-m$ CCDF is obtained in (\ref{eqCCDF00}).

\section{Mathematica$\texttrademark$ program for the computation of $\Phi_3$ function}
\label{App0}
\small
\begin{verbatim}
InvLap[f0_, x0_] := Module[{f = f0, x = x0},
   fk[eje_] := Cos[x eje];
   a = 1 + 1/x;
   fg[eje_] := (2 Exp[a x])/\[Pi] f[a + I eje];
   psum1[1] := NIntegrate[Re[fk[eje] fg[eje]], {eje, 0, \[Pi]/(2 x)}];
   psum1[i_?NumberQ] := NIntegrate[Re[fk[eje] fg[eje]], 
   {eje, -(\[Pi]/(2 x)) + i \[Pi]/x , -(\[Pi]/(2 x)) + (i + 1) \[Pi]/x}];
  psum1[1] +  NSum[psum1[i], {i, 1, \[Infinity]}, 
  Method -> "AlternatingSigns", "VerifyConvergence" -> False]]
\end{verbatim}
(* Implementation of $\Phi_3(\beta,\gamma;zt,yt)$*)
\begin{verbatim}
ConfluentHypergeometric\[CapitalPhi]3[\[Beta]\[Beta]_, \[Gamma]\
\[Gamma]_, zz_, yy_, tt_] :=
 Module[{\[Beta] = \[Beta]\[Beta], \[Gamma] = \[Gamma]\[Gamma], 
   z = zz, y = yy, t = tt, f},
   f[s_] := Gamma[\[Gamma]]/s^\[Gamma] (1 - z/s)^-\[Beta] Exp[(y)/s];
   1/t^(\[Gamma] - 1) InvLap[f, t]]
\end{verbatim}

\normalsize

\bibliographystyle{ieeetr}
\bibliography{Bivariate}

\clearpage

\clearpage

\clearpage

\clearpage

\clearpage

\clearpage


\end{document}